\documentclass[aps,reprint,onecolumn,notitlepage]{revtex4-1}
\usepackage{amsmath, amssymb, color, enumitem, graphicx}

\usepackage{natbib}
\usepackage{amsmath,amsbsy,amssymb,amsfonts}

\usepackage{color}
\usepackage{bm}
\usepackage{graphicx}
\usepackage{xfrac,nicefrac}
\usepackage{url}
\usepackage{cleveref}
\usepackage{epstopdf, epsfig}

\newcommand{\parD}[2]{\frac{\partial #1}{\partial #2}}

\renewcommand{\vector}[1]{\mathbf{#1}}
\renewcommand{\tensor}[1]{\mathbb{#1}}

\newcommand{\vv}{\vector{v}}
\newcommand{\vu}{\vector{u}}
\newcommand{\vn}{\vector{n}}
\newcommand{\vx}{\vector{x}}
\newcommand{\domain}{\mathcal{D}}
\newcommand{\surface}{\mathcal{S}}
\newcommand{\traction}{\bm{\tau}}

\newcommand{\vint}[1]{\int_{\domain} #1 \, dS}

\newcommand{\Rey}{Re}
\newcommand{\Pen}{Pe}

\newif\ifshowcomments
\showcommentstrue

\ifshowcomments
\newcommand{\PES}[1]{{\color{red}{\em #1}}}
\newcommand{\KSH}[1]{{\color{blue}{\em #1}}}
\else
\newcommand{\PES}[1]{}
\newcommand{\KSH}[1]{}
\fi

\usepackage{soul}

\begin{document}
\title{Axisymmetric Squirmers in Stokes Fluid with Non-Uniform Viscosity}
\author{Patrick S. Eastham}
\affiliation{Department of Mathematics, Florida State University,
Tallahassee, FL 32304, USA}
\author{Kourosh Shoele}
\affiliation{Department of Mechanical Engineering, Florida State University,
Tallahassee, FL 32310, USA}

\begin{abstract}
The ciliary locomotion and feeding of an axisymmetric micro-swimmer in a complex fluid whose viscosity depends on nutrient concentration are investigated numerically.  The micro-swimmer is modeled as having spheroidal geometry, and ciliary beating is modeled by a slip velocity; i.\,e. a squirmer is adapted. 
Looking at the coupling between swimming and feeding of spheroidal squirmers, it is found that swimming speed and feeding are most affected by a non-uniform viscosity environment when the ratio of advection-forces to diffusion-transport, characterized by the nondimensional Pecl\'et number, is moderate ($\Pen{}\approx 10$). These changes are correlated to significant increases in the pressure force on the surface of the squirmer. The swimming and feeding changes are found to be more significant in oblate spheroids than prolate spheroids. Most interestingly, nontrivial symmetric slip boundary conditions on the surface of the squirmer, which results in zero net motion in a constant viscosity fluid, can yield non-zero propulsion when paired with asymmetric nutrient boundary conditions. The fact that symmetric fluid boundary conditions can result in asymmetric propulsion has implications for the design of artificial micro-swimmers and other low-\Rey{} squirmers in complex fluids. 
\end{abstract}

\maketitle

\section{Introduction}
Studying the locomotion principles of micrometer-scale organisms -- commonly called micro-swimmers -- inspires both biologists and engineers who wish to develop artificial micro-robots for numerous applications, such as targeted drug delivery \citep{darnton2004moving,zhuang2017propulsion,schauer2018motility,cho2012development,zheng2016motility} and miniature noninvasive surgery \citep{nelson2010microrobots,peyer2013bio}. For efficient nutrient acquisition and colonization, many microorganisms adopt a particular means of swimming through evolutionary pressures. Micro-swimmers such as nematodes, bacteria, protozoa, and algae usually use appendages to move through their environment. These appendages can be categorized into two groups based on their morphology. The first group, flagella, are long slender appendages and are small in number \citep{taylor1951analysis,sauzade2011taylor,cortez2018regularized,berg1973bacteria}. The second group, cilia, are shorter relative to swimmer body size and often exist in groups of thousands \citep{lodish2016}. This paper employs a common mathematical model of ciliated swimmers -- the squirmer model -- in a complex fluid environment where the fluid viscosity is pointwise dependent on a surrounding scalar field. Here the scalar field is interpreted as a nutrient; however, without loss of generality, it can be interpreted as other physical quantities such as temperature or other chemicals.

An example of micro-swimmers in a complex fluid is volvocine green algae. This algae starts from a simple biflagellate cell, \emph{Chlamydomonas}, and can organize to form regularly-shaped species with up to billions of packed \emph{Chlamydomonas}-like cells \citep{herron2009triassic}. A particular member of this family is Volvox, a spheroidally-shaped species up to several millimeters in diameter with thousands of moving flagella on its surface. A multi-cellular Volvox could have a large rate of nutrient uptake and waste disposal and lives in a high-Pecl\'et number flow regime \citep{short2006flows,solari2006multicellularity,goldstein2015green} where the nutrient field exhibits a flow-induced asymmetry near the body. Colonies of Volvox also exhibit enhanced mixing with a non-Gaussian distribution of tracer displacement \citep{kurtuldu2011enhancement,zaid2011levy}. It was proposed by \citet{lin2011stirring} that the near-field velocity field and the existence of stagnation point in the flow are responsible for the emergence and extent of non-Gaussian statistics. Previous theoretical and experimental work on the diffusiophoretic motion of passive Janus spheres also shows the presence of such non-Gaussian behavior which is traced back to the presence of the superdiffusive regime next to the passive swimmer with pro-dominant deterministic particle displacements \citep{zheng2013non}. The question is whether nutrient-dependent flow characteristics and induced material non-linearity can influence the flow statistics near the body and contribute to the deviation from Gaussian statistics in this system. It might also constructively influence the rotational steering observed in Volvox carteri \citep{hoops1999test,bennett2015steering} and \emph{C. crescentus} \citep{morse2013molecular}. Another example is the bacterium \emph{H. pylori}. This bacteria lives in the human stomach and uses enzymes to transform impenetrable stomach mucus into a softer material in which the organism has greater motility \citep{celli2009helicobacter}. This modification of the environment by the microorganism has an important impact on the surprising success of \emph{H. pylori} to infect humans, which by recent estimates is 50\% of the world’s population \citep{ottemann2002helicobacter}. 

One of the fundamental needs of bacteria is to search for new sources of nutrient and, when found, to feed on these sources. This raises a question of whether it is possible for an organism to use its feeding mechanism to enhance locomotion towards nutrient-rich regions. Another motivation of this study is based on recent findings about the ciliary locomotion modes in micro-swimmers.  For ciliary locomotion, a common simplification of the model is to assume a steady, tangential velocity at the bacteria surface, first proposed by \citet{blake1971spherical}. This type of motion is commonly known as the squirmer model. \citet{michelin2011optimal} found that, for the constant-viscosity case, the optimal swimming and optimal feeding stroke are both equivalents to ``pure treadmill" motion. This is interesting because it suggests that the other stroke modes, which produce mixing effects but no new swimming propulsion, are unnecessary for these important biological functions. We will examine how the coupling between nutrient and viscosity affects swimming and feeding and whether the same conclusion can be reached for a nutrient-dependent viscous fluid.

Finally, the coupling between scalar transport and flow dynamics can be leveraged to manufacture more efficient swimming micro-robots.  For example, squirmer motion has been used in micro-robotic applications \citep{huang2019adaptive,cho2014mini,li2017micro}, wherein two modes of swimming have been employed: passive swimming and active swimming. Inert particles generate passive swimming through the use of gradients in a surrounding scalar field. These gradients generate a slip-velocity on the surface of the particle and results in phoretic propulsion. In the case of active swimmers, the slip velocity on the surface of the particle is generated by the particle itself (e.g., micro-swimmers, bacteria, etc). It is interesting to find whether an active swimmer can benefit from the passive mechanisms of well-coordinated scalar transport. In particular, this paper considers a combination of passive and active effects; the squirmer is an active swimmer whose cilia produce a slip velocity on the surface, but the dependence of the fluid viscosity on the surrounding nutrient concentration allows a feedback mechanism for the stroke motion to produce passive phoretic propulsion. Applications of findings in this paper are not limited to living bacteria, as variable-viscosity can occur in more general cases, such as when there is heat dissipation into surrounding fluids from an immersed body. For instance, the viscosity of water changes by order of magnitude between freezing and boiling \citep{korson1969viscosity}, so the effects of variable viscosity for micro-robots who use thermal propulsion such as Janus-like particles \citep{jiang2010active} can benefit from the results presented in this paper. Motivated by these biological and micro-robotic applications, the strong feedback effect between a surrounding scalar field and flow dynamics around a canonical free-swimmer model is investigated in this paper.

When the environments under consideration are governed by linear, constant-coefficient PDEs, as is the case in constant-viscosity Stokes equation, one can employ classical techniques to find fundamental solutions of the flow field produced by ciliary beating \citep{pak2014generalized,nganguia2018squirming}. This approach has been employed to study ciliary micro-swimmers in Newtonian fluids \citep{elgeti2015physics,lighthill1952squirming,blake1971spherical,keller1977porous,lauga2012microswimmers,michelin2011optimal,berg1973bacteria,pak2014generalized,pedley2016spherical}.
More recently, there has been an interest in understanding how the swimming performance of micro-swimmers changes in a complex fluid environment with nonlinear rheology. This is motivated by the fact that real-world fluid environments contain numerous complexities such as proteins,  particles, polymers, macromolecules, and temperature \citep{patteson2016active}. Any change in these components can induce local modification to the rheology of the fluid medium and consequently affect the governing constitutive relation between stress and shear rate. 

When the governing equations are only weakly nonlinear, asymptotic techniques can be used to obtain information about an object's propulsion \citep{morozov2019self}. These asymptotic techniques can be accompanied by the powerful Generalized Reciprocal Theorem \citep[pp. 85-87]{happel1973low} to explain swimming performance changes \citep{pozrikidis2016reciprocal,elfring2015note}. This technique --  where a known auxiliary solution is employed to construct asymptotic corrections to quantities such as swimming speed -- has been used to study the weakly-coupled effect of shear-thinning viscosity or viscoelasticity on micro-swimmer mobility \citep{datt2015squirming,nganguia2017swimming,datt2017active}. The authors used this technique in their previous paper to study the effects of a weak dependence of viscosity on a nutrient field surrounding a squirmer \citep{shoele2018effects}. It was shown that, when the fluid has a nutrient-dependent viscosity, the swimming performance changes across different modes of swimming. For a weak dependence, it was observed that the optimal swimming speed of the micro-swimmer could further increase due to the spatial variation of the fluid viscosity. This suggests that a micro-swimmer could harness the viscosity-modifying properties of nutrient concentration to enhance its locomotion. However, since the effect of variable viscosity on nutrient flux occurs at higher-order terms than examined, the analytic approach employed before cannot be used to examine the effect of variable-viscosity on nutrient uptake by the swimmer.

When there is substantial fluid nonlinearity as a result of viscosity changes (or there is ``strong coupling" between the viscosity changes and fluid flow), there is no analytical technique available, and therefore numerical simulations are required to study the system. Complex fluids such as shear-thinning or viscoelastic fluids, where the effective viscosity of the fluid is modified as a result of viscoelasticity and swimming gaits by the micro-swimmer, have been studied \citep{datt2015squirming,nganguia2018squirming,nganguia2017swimming}. This paper, however, explores the novel situation of a strong dependence between viscosity and a surrounding scalar field transported by the fluid (e.g., nutrient, chemicals or heat) and investigates how the spatially varying viscosity affects the propulsion performance and nutrient uptake of spheroidal  squirmers.  This is potentially important for microorganisms with high Pecl\'et number waste disposal or nutrient uptake and for artificial micro-swimmers who passively modify their surrounding chemical or temperature field to induce phoretic locomotion.

To solve this problem -- a coupled Stokes and Advection-Diffusion system -- the Finite Element Method is employed. The generality of this method allows us to examine complex domains that would be difficult to achieve with other methods such as Finite Differences \citep{strikwerda2004finite}. Although boundary integral \citep{smith2009boundary,cortez2001method} and immersed boundary \citep{peskin2002immersed} methods have been employed to study the beating of flagella when cilia are fully resolved; the squirmer model used in this study allows for the approximation of the beating motion of cilia using tangential velocity boundary conditions to explore the problem efficiently. The spheroidal squirmer canonical model allows us to explore the feedback mechanisms between passive and active swimming methods and their relation with the geometry of the swimmer in a computationally efficient manner.

This paper proceeds as follows: section \ref{sec:model} describes the squirmer model, first as a general description and then using specific squirmer assumptions and its numerical discretization, section \ref{sec:results} contains detailed descriptions of the results for different parameter regimes. Section \ref{sec:conclusion} discusses  the importance and significance of our findings.

\section{Nutrient Transport around a Squirmer} \label{sec:model}
A spheroidal squirmer with surface $\surface$ is considered in an infinite domain $\domain$ at negligible Reynolds number $\Rey{} =\rho UL/\mu_0\ll0$ where $\rho$ is the fluid mass density, $U$ is the characteristic swimming velocity, $L$ is the diameter of the squirmer, and $\mu_0$ is the characteristic (far-field) viscosity of the fluid. For the case of nutrient-dependent viscosity, both  free-swimming velocity and nutrient flux of the micro-swimmer depend on the Pecl\'et number of the nutrient field, the functional relationship between nutrient and viscosity, the squirmer motion, the shape of the swimmer, and nutrient absorption boundary conditions. All of these effects are discussed in section \ref{sec:results}. The squirmer model in a spatially varying nutrient field is first described in a general form in section \ref{sec:generalmodel} and then simplified for our specific axisymmetric squirmer configuration in section \ref{sec:specificmodel}. We then describe our method for numerical discretization of the strongly-coupled system by the Finite Element method in section \ref{sec:femmodel}.

\subsection{General Squirmer in a Nutrient-Dependent Fluid} \label{sec:generalmodel}

We follow the general model previously proposed in \citet{michelin2011optimal} and briefly restated here for convenience. In a body-fixed coordinate system, the fluid is governed by Stokes equations
\begin{subequations} \label{eq:stokes}
\begin{eqnarray}
\nabla\cdot\tensor{T}  = 0,&\qquad \mbox{for\ } \vx\in\domain \\
\nabla\cdot\vector{u} = 0,&\qquad \mbox{for\ }\vx\in\domain \\
\vu = \vu^\surface,&\qquad \mbox{for\ } \vx\in\surface \\
\vu \to -(\vector{U}+\bm{\Omega}\times\vx)&\qquad \mbox{as\ } ||\vx|| \to \infty \label{eq:farfieldvel}
\end{eqnarray}
\end{subequations}
where $\vector{u}$ is the flow velocity vector, $\tensor{T}=-p\tensor{I}+2\mu\tensor{D}$ is the viscous stress tensor with the spatially varying viscosity $\mu=\mu(\vx)$, $p$ is the fluid pressure, $\tensor{D} = \frac{1}{2}(\nabla\vector{u} + \nabla\vector{u}^\top)$ is the {strain rate tensor}, $\vector{u}^\surface$ is the squirmer motion along the surface and $\vector{U}$ and $\bm{\Omega}$ are the translational and rotational velocities of the squirmer, respectively. 

The squirmer is assumed to be a free swimmer at zero Reynolds number where inertial effects are negligible. This implies the free-swimming condition which corresponds to no drag and torque on the body of the squirmer,
\begin{equation}
\int_{\surface} \traction \, dS = \vector{0},\qquad \int_{\surface} \vx \times \traction \,dS = \vector{0}  \label{eq:drag}
\end{equation}
where $\traction = \vector{n}\cdot\tensor{T}$ is the traction on the surface $\surface$. 

The equations are non-dimensionalized as follows. Characteristic length $\alpha$ is chosen to be the square root of the surface area of the squirmer, which is fixed, and characteristic velocity is $\sqrt{\mathcal{P}_0/\mu_0\alpha}$, where $\mathcal{P}_0$ is the power expenditure by an equivalent spherical squirmer (with the same surface area) in a constant-viscosity fluid, defined as 
\begin{equation}
\mathcal{P} =  \int_\domain \tensor{T}:\tensor{D}\, dV 
= -\int_\surface \vu^\surface\cdot(\vn\cdot\tensor{T})\, dS. 
\label{eq:power} 
\end{equation} 

This non-dimensionalization is chosen so that our system can be applied to -- and comparisons can be made between -- different spheroidal geometries. 

The scalar nutrient $C$ is fully absorbed on a portion of (and possibly all of) the surface $\surface_1$ of the swimmer and approaches a limiting value of $C_\infty$ in the far-field. For convenience, we normalize the nutrient as  $\displaystyle c=(C_\infty - C)/{C_\infty}$ and solve the advection-diffusion equation
\begin{subequations} \label{eq:advdiff}
\begin{align}
\Pen\,\big[\vector{u}\cdot \nabla c\big] = \nabla^2 c,&\qquad \mbox{for\ } \vx\in\domain \\
c = 1,&\qquad \mbox{for\ } \vx\in\surface_1 \\
\parD{c}{n}=0,&\qquad \mbox{for\ } \vx\in\surface_2 \\
c\to 0,&\qquad \mbox{as\ } ||\vector{x}||\to\infty
\end{align}
\end{subequations}
where nutrient is completely absorbed on $\surface_1$ and there is no nutrient flux on $\surface_2$; formally we require $\surface_1\cup\surface_2=\surface$ and $\surface_1\cap\surface_2=\varnothing$. The Pecl\'et number is defined using the previously mentioned characteristic length and velocity
\begin{equation}
\Pen{}= \frac{1}{\kappa}\sqrt{\frac{\mathcal{P}_0\alpha}{\mu_0}} 
\end{equation}
with $\kappa$ being the nutrient diffusivity. Through a suitable rescaling, the normalized scalar field $c$ can be considered as a different physical quantity, such as temperature or a chemical agent \emph{emitted}, as opposed to absorbed, by the squirmer. Through equation \ref{eq:viscnutrient}, the viscosity depends on the nutrient field, and the nutrient field depends on the viscosity (through the velocity field), so equations \ref{eq:stokes} and \ref{eq:advdiff} have a two-way coupling.

\begin{figure}
\centering
\includegraphics[scale=1]{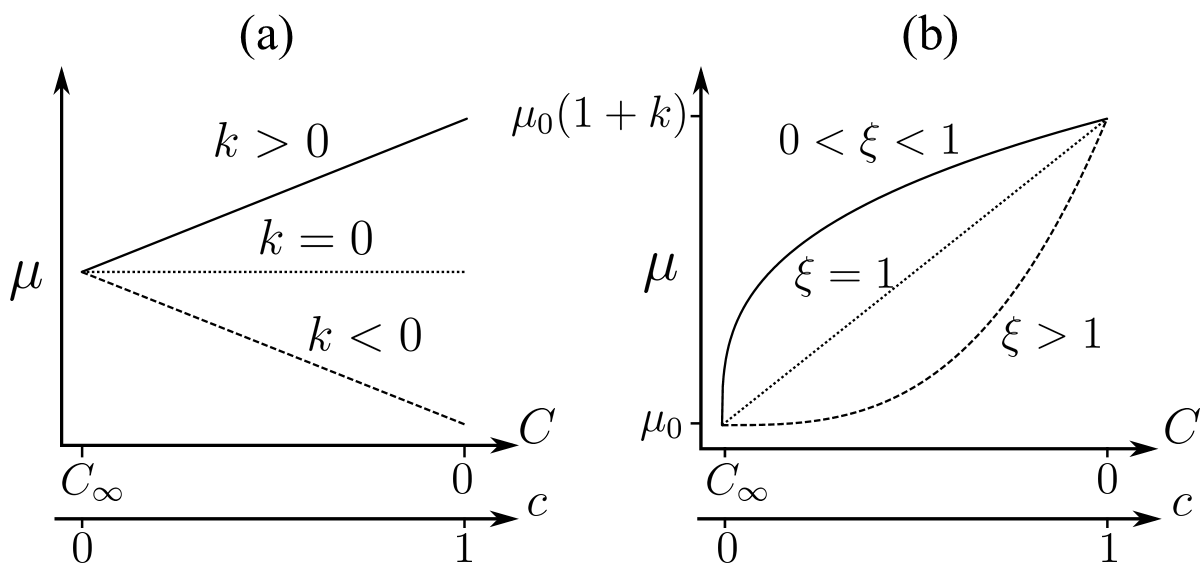}
\caption{The nutrient-viscosity relationship for (a) various $k$, fixed $\xi=1$, and (b) various $\xi$, fixed $k>0$, based on equation \ref{eq:viscnutrient}. The $\xi$ determines the type of relationship, while the magnitude and sign of $k$ determines whether the viscosity increases or decreases near the surface of the squirmer ($C=0$, $c=1$). Note that $C$ is the dimensional nutrient variable and $c$ is its dimensionless counterpart.}
\label{fig:NutrVisc}
\end{figure}

Feeding by the micro-swimmer is mathematically modeled as the nutrient flux at the surface $\surface$ and is defined in non-dimensional form as 
\begin{equation}
\Phi = -\frac{1}{\Pen}\int_\surface\parD{c}{n} dS 
\label{eq:nutrientflux}
\end{equation}
where $\partial c/\partial n=\vn\cdot\nabla c$ and $\vn$ is the unit normal vector on $\surface$ pointing into the flow. We follow \citet{shoele2018effects} in defining the canonical pointwise dependence of viscosity on nutrient
\begin{equation}
\mu\big(c(\vx)\big) = \mu_0\left[1 + k c(\vx)^\xi\right] \label{eq:viscnutrient}
\end{equation}
where $k$ and $\xi$ are parameters that govern the strength and qualitative nature of the relationship, respectively. The characteristic viscosity $\mu_0$ corresponds to the viscosity of the nutrient-saturated fluid, i.e.\ the viscosity in the far-field. Qualitatively, $k<0$ corresponds to viscosity increasing with increasing nutrient $C$, as is the case with sugars, nitrate, and other complex proteins. On the other hand, $k>0$ corresponds to viscosity increasing with decreasing $C$, for example when $C$ is used to model a temperature field or a different chemical whose presence decreases fluid viscosity \citep{telis2007viscosity,wiesenburg1988synopsis,kodejvs1976viscosity,strauss1953polyphosphates,mcbride2014viscosity}. The power $\xi$ is analogous to those used in power-laws for shear-dependent viscosity flows, employed here to represent the potentially nonlinear relationship between nutrient concentration and viscosity. Figure \ref{fig:NutrVisc} illustrates the qualitative behavior of $\mu$ with respect to nutrient based on equation \ref{eq:viscnutrient}.

\subsection{Steady Squirmer Model for a Spheroid} \label{sec:specificmodel}
\begin{figure}
 	\centering
	\includegraphics[scale=0.26]{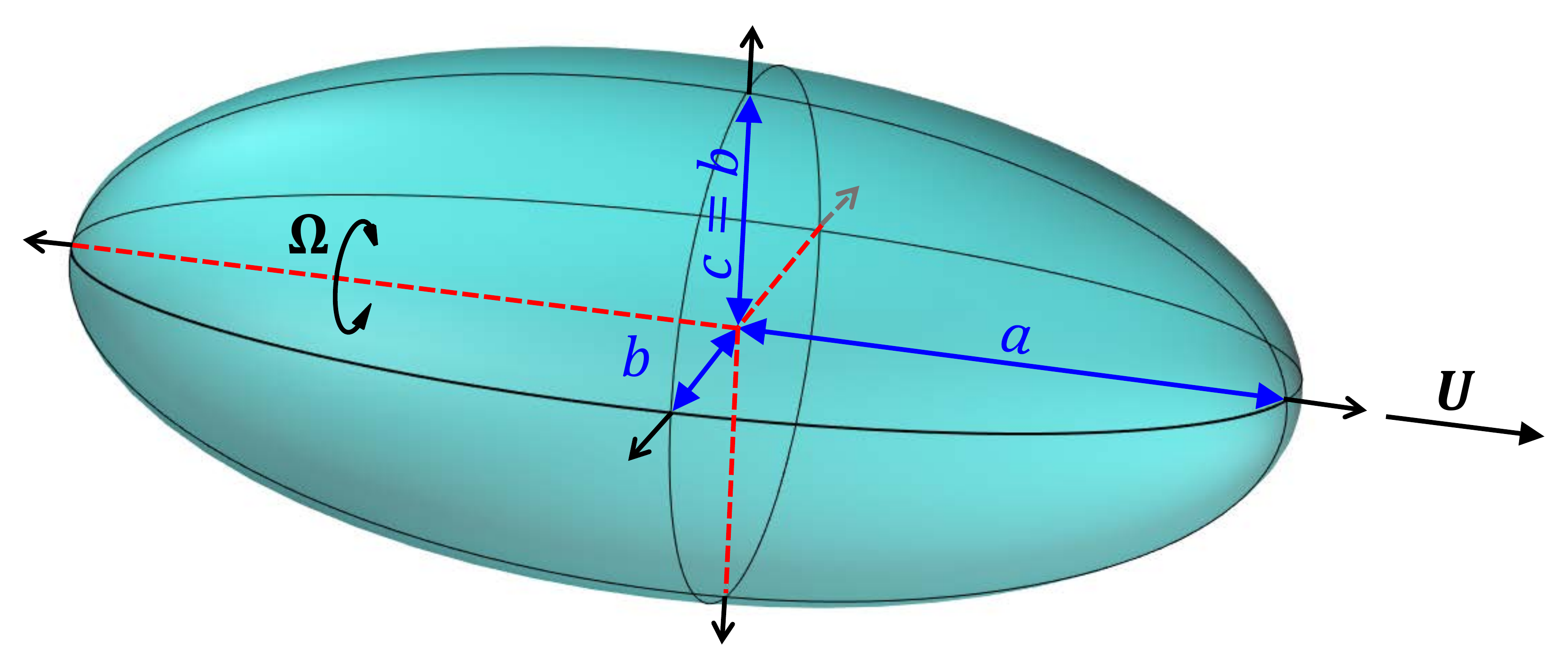} 
	\caption{Spheroidal micro-swimmer where $c=b$. It is called prolate if $a>b$, spherical if $a=b$, and oblate if $a<b$. The squirmer model corresponds to $\bm{\Omega}=\vector{0}$ and $\vector{U}$ which satisfies the zero-drag condition.}
	\label{fig:ellipsoid}
\end{figure}

Spheroidal geometries are common across many micro-swimmers \citep{young2006selective} and is used here as a canonical geometry. The squirmer surface $\mathcal{S}$ is chosen to be a spheroid, also known as an axisymmetric ellipsoid, with principal semi-axes $a$, $b$ and $c$ where $b=c$ provides axisymmetry (see figure \ref{fig:ellipsoid}). We will characterize our spheroid by the aspect ratio $\ell=a/b$ where for $\ell<1$ or $\ell>1$ the surface $\mathcal{S}$ is called an oblate or prolate spheroid, respectively. As mentioned previously, in order to compare the swimming and feeding of  various spheroids, we fix the surface area of the swimmer, $\alpha^2$, for all aspect ratios $\ell$. This is motivated by the observation that the power expenditure and feeding rates of a micro-swimmer are proportional to its surface area. A squirmer with the same parameters in a constant-viscosity fluid is used as the benchmark system for comparing the effect of fluid with nutrient-dependent viscosity on swimming and feeding. 

The surface area $\alpha^2$ can be approximately computed based on the semi-axes of the spheroid with 
\begin{equation}
\alpha^2 = 4\pi\left(\frac{(ab)^p + (ac)^p + (bc)^p}{3}\right)^{1/p} 
\label{eq:S_function}
\end{equation}
where for $p=1.6075$ the error is bounded by $\pm 1.061\%$ \citep{Thomsen2014}. For a given aspect ratio, we need to choose $a$ and $b$ to keep the surface area constant. We let $b=c$ and solve equation \ref{eq:S_function} for $b$,
\begin{equation}
b = b(\ell) = \left(\frac{3\left(\frac{\alpha^2}{4\pi}\right)^p}{2\ell^{p} + 1}\right)^{1/(2p)}.
\label{eq:b_function}
\end{equation}

The surface motion consists of a decomposition into orthogonal ``stroke modes'' and the tangential surface velocity $\vu^\surface = u_t^\surface\vector{e}_t$, where $\vector{e}_t$ is the tangential vector on the surface of the spheroid, is expressed as
\begin{equation}
u_t^\surface(\gamma) = \sum\limits_{n=1}^\infty \beta_n K_n(\gamma)
\label{eq:surfacestroke} 
\end{equation}
where for each orthogonal mode $n$, $K_n(\gamma)$ is defined as $K_n(\gamma) = \sqrt{\frac{3}{n(n+1)}}\sqrt{1-\gamma^2}P_n'(\gamma)$ with $P_n$ being the $n$th-order Legendre polynomial and $\gamma=\cos(2\pi\chi/L_E)$. Here, $L_E$ is the total arclength of the ellipse with semi-axis $a$ and $b$ and $\chi$ is the arclength from the front point $(a,0,0)$ to a particular point on the surface. We note that $\vu^\surface$ is slightly modified from \citet{shoele2018effects}, where only spherical shape was considered, to adjust for arc-length along the spheroid. 

These surface stroke modes $\beta_n$ are constrained by 
\begin{equation}
\frac{2}{3}\beta_1^2 + \sum\limits_{n=2}^\infty \beta_n^2 = 1 
\label{eq:beta_constraint}
\end{equation}
which, in the constant-viscosity case, fixes the power expenditure of the squirmer.


This paper addresses the following question: for a complex fluid whose viscosity depends on a surrounding scalar field, how do the free-swimming velocity and diffusive flux of the swimmer compare to the corresponding constant-viscosity velocity and flux? More specifically, are there any notable improvements that the complex fluid provides to the micro-swimmer over the constant-viscosity fluid? As the two-way coupling introduces a strong nonlinearity into our governing equations, answering this question requires numerical simulation.


\subsection{Numerical Methodology} \label{sec:femmodel}
We employ the Finite Element method \citep{elman2014finite,johnson2012numerical} to numerically calculate the flow and nutrient fields around the free swimming, steady squirmer. The model is written in the Julia programming language \citep{bezanson2017julia} and uses mesh data files generated in GMSH \citep{geuzaine2009gmsh}.  We follow the derivation given by \citet{tabata1996finite} for the axisymmetric Stokes equation and reformulate equation \ref{eq:stokes} into the following weak form:
\begin{subequations} \label{eq:FEM}
\begin{align}
\vint{2\mu\left[\tensor{D}(\vu):\tensor{D}(\vv) + \frac{u_yv_y}{y^2}\right]&y} - \vint{p\nabla\cdot \vv}= 0, 
\label{eq:weakForm_U_PART1} \\
\vint{q\nabla\cdot \vu}&= 0, 
\label{eq:weakForm_U_PART2}
\end{align}
\end{subequations}
for proper test functions $q$ and $\vv=v_x\hat{\vector{e}}_x + v_y\hat{\vector{e}}_y$, where $\domain$ is  a cross-section through our 3D domain along the axis of symmetry. Note that $x$ is the axis of symmetry and $y\geq0$ is the radial coordinate. The axisymmetric weak form of the advection-diffusion system (equation \ref{eq:advdiff}) is
\begin{equation}
\Pen{}\vint{(\vu\cdot\nabla c)\psi} = \vint{\left[\nabla c\cdot\nabla \psi - \frac{1}{y}\parD{c}{y}\psi\right]} 
\label{eq:weakForm_C_PART1}
\end{equation}
where $\psi$ is a test function. The flow solution $(\vu,p)$ is approximated by quadrilateral $Q_2-Q_1$ Taylor-Hood elements and $c$ is approximated by a $Q_2$ element. The model is carefully validated for different canonical problems and closed-form solutions. The velocity and scalar $c$ converge with $O(h^{-4})$, consistent with $Q_2$ elements, and pressure converges with $O(h^{-2})$, consistent with $Q_1$ elements, where $h$ is the characteristic edge length for the mesh. Based on the convergence study, the mesh resolution of approximately $h\approx0.02$ was chosen near the squirmer surface and extended exponentially to $||\vx||=20,000$. This mesh gives us approximately 4 digits of accuracy. The $h$ at the surface is chosen to encapsulate the boundary layer fully at the highest \Pen{} simulated.

To solve the coupled system \ref{eq:FEM} and \ref{eq:weakForm_C_PART1}, we employ an iteration scheme;we solve one system, then use that solution as a parameter in the other system. Specifically, the Stokes system \ref{eq:FEM} is solved first, then the velocity field $\vu$ is used in \ref{eq:weakForm_C_PART1}. The resulting $c$ field is used to compute the $\mu$ field in \ref{eq:FEM} using equation \ref{eq:viscnutrient}, then we solve the stokes system again, etc. A solution $\{\vu,p,c\}$ to equations \ref{eq:FEM} and \ref{eq:weakForm_C_PART1} is said to have converged only when a difference tolerance of $10^{-6}$ is achieved for all discretized variables.

\section{Results} \label{sec:results}
To understand how nutrient-dependent viscosity affects squirmer's performance, we compute two normalized parameters, $V$ and $J$, to quantify the change in a squirmer's free-swimming velocity and nutrient flux, respectively, between variable- and constant-viscosity fluids. 

The normalized speed modification $V$ is defined as $V=(U-U_0)/U_0$ where $U$ is the free-swimming velocity of the variable-viscosity spheroid and $U_0$ is the free-swimming velocity of the same spheroid in a constant-viscosity fluid, both defined based on equations \ref{eq:farfieldvel} and \ref{eq:drag}. The $V>0$ cases correspond to the improvement of the swimming performance of the swimmer in the variable-viscosity environment. Similarly, the normalized flux modification $J$ is defined as $J=(\Phi-\Phi_0)/\Phi_0$, where $\Phi$ and $\Phi_0$ are the diffusive nutrient fluxes of squirmer in a variable- and constant-viscosity fluid, respectively, as defined in equation \ref{eq:nutrientflux}. Regimes with $J>0$, and $c$ being interpreted as nutrient concentration correspond to feeding enhancement in the variable-viscosity environment.


In section \ref{sec:r1} we will examine how $V$ and $J$ depend on \Pen{} for a spherical treadmill squirmer at selected representative values of $k$. Then in section \ref{sec:r2} we explore $\xi$-$k$ space to determine how the nutrient-dependent viscosity relationship affects performance.  Section \ref{sec:r3} is about how the coupling between the surface modes affect performance and, in section \ref{sec:r4}, the effect of aspect ratio of an spheroid is studied. For sections \ref{sec:r1} to \ref{sec:r4} $\surface_2=\varnothing$, indicating nutrient being fully absorbed on the entire surface. Finally in section \ref{sec:r5} we take a close look at how asymmetric $c$ distribution on the swimmer surface could generate non-zero propulsion. This asymmetric $c$ distribution is generated by letting $\surface_2\neq\varnothing$ such that nutrient is being absorbed on only a part of the surface.

\subsection{Spherical Treadmill Squirmer} \label{sec:r1}
\begin{figure}
    \centering
    \includegraphics[scale=0.8]{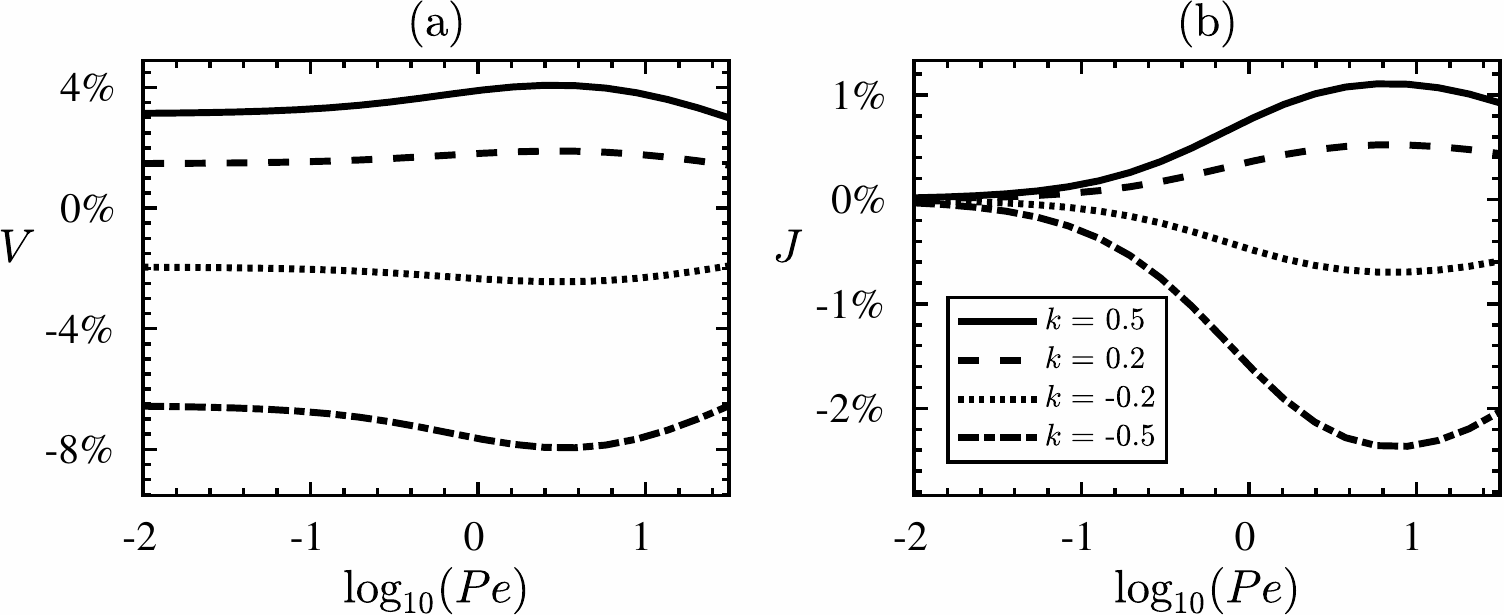}
    \caption{Effect of variable viscosity on a spherical treadmill squirmer with $\xi=1$. (a) $V$ increases monotonically with $k$, is non-zero even for diffusion-domainted regimes and shows extrema for $\Pen\approx 10$ (b) $V$ also increases monotonically with $k$, is zero for diffusion dominated regimes, and exhibits extrema around the same $\Pen$ as $V$.}
    \label{fig:r1_params}
\end{figure}

In this section, a spherical treadmill squirmer ($\beta_1=\sqrt{3/2}$) for fixed $\xi=1$, i.e. linear dependence of viscosity on $c$, is chosen; we see the effect of variable viscosity on $V$ and $J$ in figure \ref{fig:r1_params}. Even for very low \Pen, $V$ is nonzero and increases monotonically with $k$; however, $V$ is \emph{not} monotonic in \Pen, and experiences extrema for $\Pen\approx 7$. In a low-\Pen{} environment $J$, unlike $V$, is unaffected by variable viscosity. As \Pen{} increases,  $J$ also show non-monotonic behavior with extreme values observed for $\Pen{} \approx 10$. The \Pen{} that produces extreme values of $V$ and $J$, while close, are not equal. 

These two observations suggest the following physical explanation: For low-\Pen, diffusion dominates and so there is no nutrient boundary layer near the surface of the swimmer. This lack of a boundary layer means that nutrient in the near-field is similar in both the constant- and variable-viscosity fluids, meaning the diffusive flux across the boundary are similar as well, as we observe. However, the drag on the surface of the squirmer is a non-local effect, which allows for a modified free-swimming velocity even without a boundary layer. As \Pen{} increases, the advection effect dominates over diffusion and a boundary layer develops. This nutrient boundary layer causes an asymmetric distribution of the viscosity around the squirmer. At $\Pen\approx 7$, the thickness of the boundary layer has maximum effect on propulsion, as there is enough viscosity to ``push off" against. As the \Pen{} increases even further, the boundary layer thins, and the viscosity of the fluid in the near-field decreases (increases) even further for $k>0$ ($k<0$).

\begin{figure}
    \centering
    \includegraphics[scale=0.8]{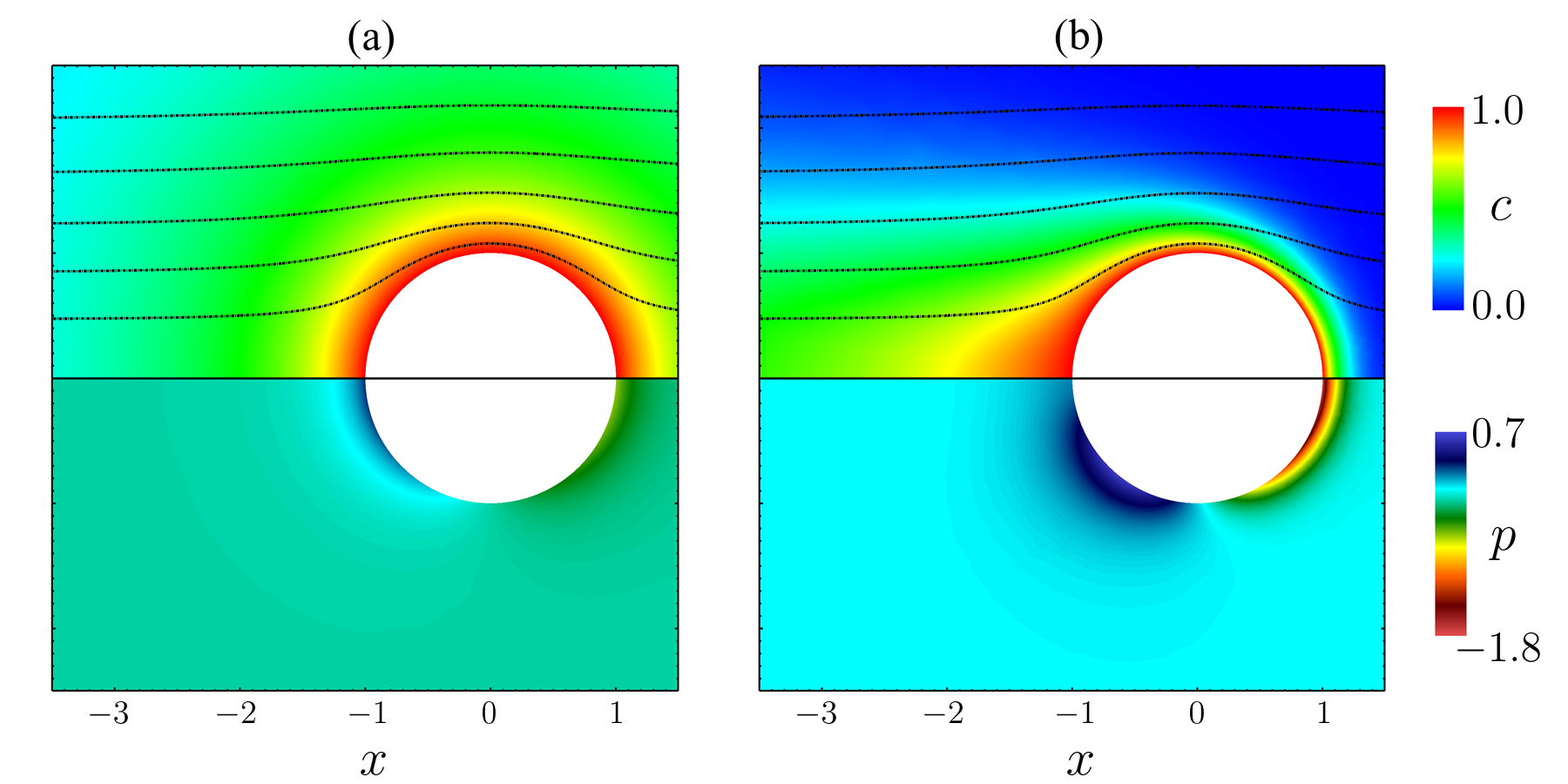}
    \caption{Effect of \Pen{} on a treadmill squirmer with $\xi=1.0$ and $k=0.5$; nutrient field with streamlines on top and pressure field on bottom. (a) $\Pen=0.01$, $V = 3.1\%$, $J=0.0\%$ both nutrient and pressure are symmetric due to the diffusion-dominated regime. (b) $\Pen=10$, $V=3.8\%$, $J=1.1\%$ advection-dominated regime causes an asymmetry in fields which results in larger $V$ and $J$.}
    \label{fig:r1_flows}
\end{figure}

Figure \ref{fig:r1_flows} shows the contour plots of $c$ and pressure of a treadmill squirmer at low and high \Pen{}. The top half is the $c$ contour field and the flow streamlines, and the bottom panel is the pressure field. Although $c$ exhibits large changes with increasing \Pen, the flow streamlines show minor modifications. The pressure field, on the other hand, undergoes a large change as \Pen{} increases. For low \Pen, the pressure is approximately symmetric about the sphere, but as \Pen{} increases, the nutrient-dependent viscosity becomes asymmetric and the total pressure difference between the front and back of the swimmer increases as well. This is the most notable observed change in the flow variables compared to the constant-viscosity squirmer. Because the streamlines are approximately similar for different \Pen{}, but pressure changes greatly, we hypothesis that variable viscosity modifies the pressure field but only induces small changes in the velocity. 

The pressure field is also found to be affected by $k$. Although not shown, the sign of the pressure field switches with the sign of $k$. Also the pressure difference across the surface of the squirmer monotonically increases with $|k|$. Because swimming performance is also monotonic in $k$, we conclude that the enhancement of the swimming performance due to variable viscosity occurs when the pressure force is in the same direction of swimming motion. While when the pressure force is in the opposite direction of the swimming, performance is hindered. Qualitatively, it appears that the velocity field is mostly affected by the boundary conditions and the incompressibility condition and shows small receptivity to the changes in the viscosity; on the other hand, the pressure gradient in the momentum equation modifies substantially to accommodate for induced changes by the spatial variation of the viscosity and in return it can induce modifications to the swimming velocity compared to the constant-viscosity case. 

\subsection{Effect of Nutrient-Viscosity Relationship} \label{sec:r2}
\begin{figure}
\centering
\includegraphics[scale=0.8]{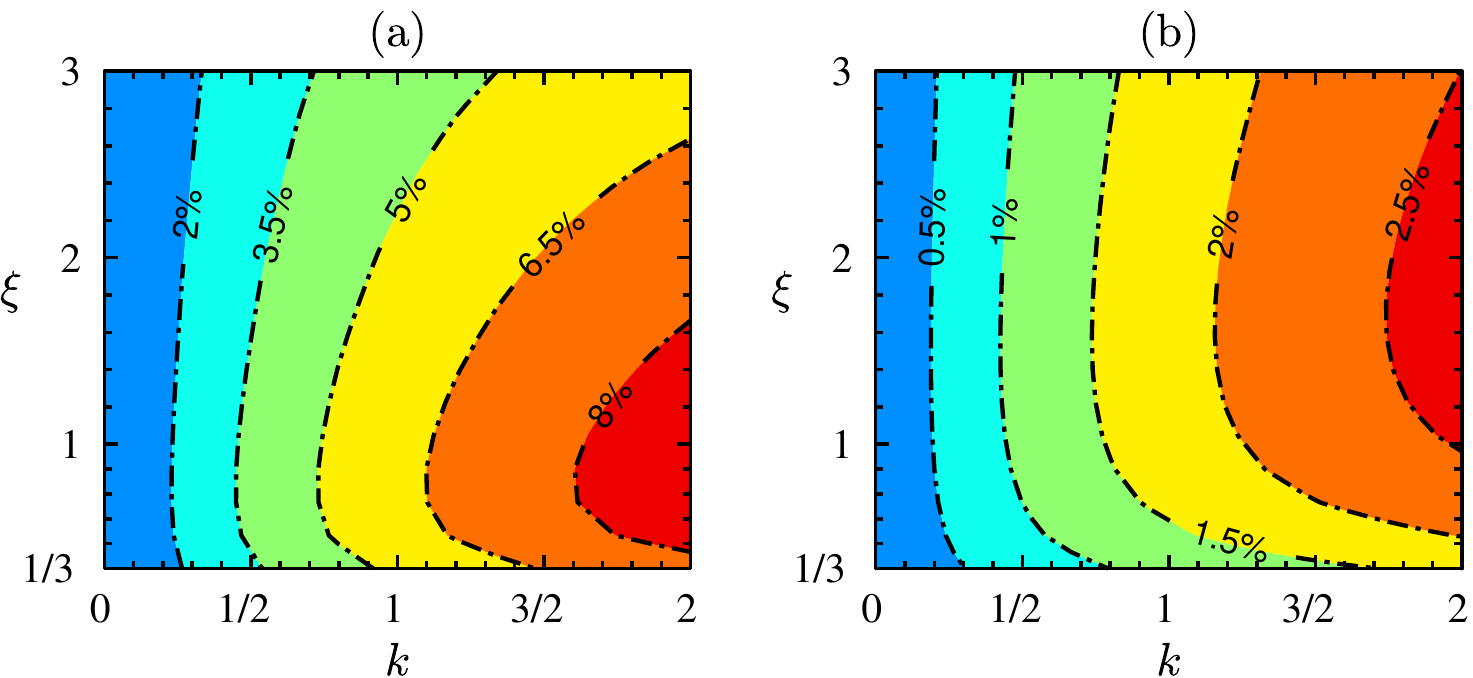}
\caption{Effect of $k$ and $\xi$ on the swimming and feeding performances. Contour plots of (a) $V$ and (b) $J$. The effect of nutrient-dependent viscosity increases monotonically with $k$ but the optimal values of $V$ and $J$ show preference for qualitatively different fluids represented with $\xi$ (linear $\xi=1$ for $V$ and nonlinear $\xi\approx 2$ for $J$).}
\label{fig:r2_params}
\end{figure}

Here, we investigate the joint effects of $\xi$ and $k$ on $V$ and $J$ of a spherical squirmer with pure treadmill motion. We fix $\Pen=10$ based on our results from section \ref{sec:r1}. Figure \ref{fig:r2_params} shows the contour plots of $V$ and $J$ for $0<k<2$ and $1/3<\xi<3$. Both $V$ and $J$ increase with $k$, but their rates of increase are dependent on $\xi$. While both $V$ and $J$ attain their maximum values at high $k$, and they increase monotonically with $k$ over all $\xi $ values, $V$ experiences a maximum for $\xi \approx 1$ while $J$ has a maximum for $\xi\approx 2$. Moreover, the region with high level of $J$ is more elongated in $\xi$ than what is present in $V$.  The current results are shown only for $k>0$ as  $k<0$ cases also show the same trends.

\begin{figure}
    \centering
    \includegraphics[scale=0.9]{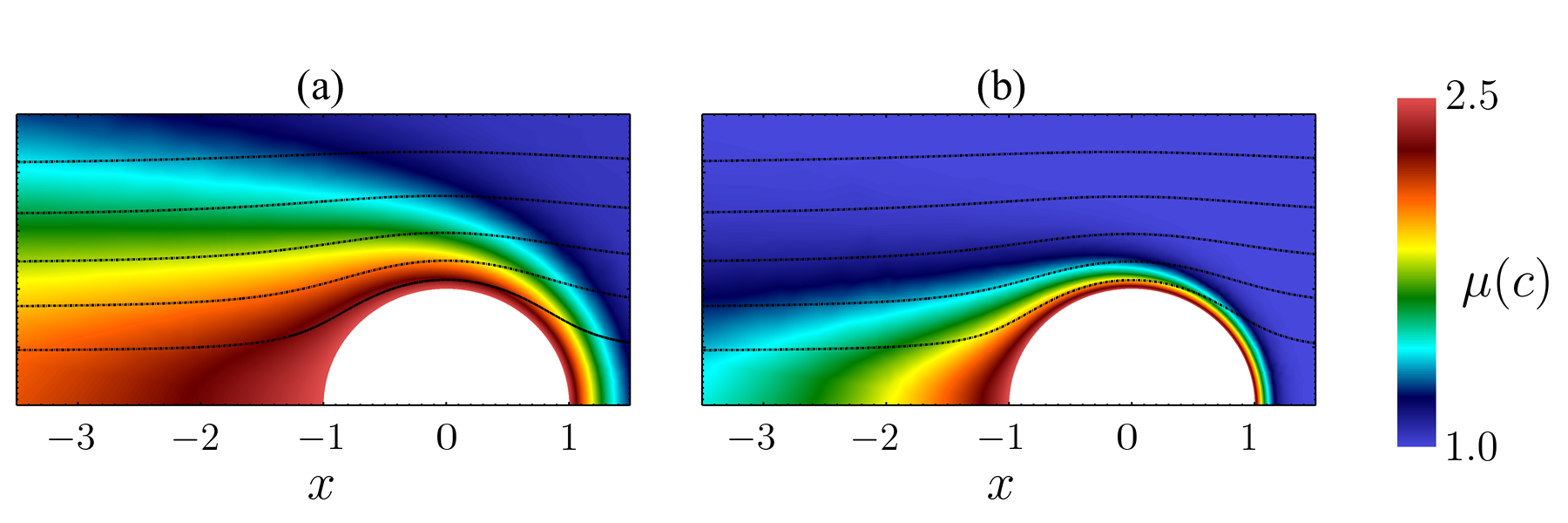}
    \caption{Variable Viscosity fields for various $\xi$ and $k$. (a) $\xi=0.5$, $k=1.5$, $V=7.4\%$, $J=1.8\%$, (b) $\xi=2.0$, $k=1.5$, $V=6.6\%$, $J=2.3\%$. Changing $\xi$ affects the $\mu$ profile which determines the thickness of the viscosity boundary layer.}
    \label{fig:r2_flows}
\end{figure}

In figure \ref{fig:r2_flows} two representative cases for the viscosity field $\mu$ for different $\xi$ are shown. We immediately notice that the streamlines are approximately the same with minor changes near the body. Although not shown here, the increasing $V$ is paired with an increase in the asymmetrical pattern and magnitude of the pressure field, as explained previously in section \ref{sec:r1}. We also notice that the viscosity field, due to its dependence on the nutrient field, depends qualitatively on $\xi$. This also confirms our initial intuition to use a power-law equation, equation \ref{eq:viscnutrient}, as a canonical model to represent the relation between the nutrient and viscosity field; by changing $\xi$, one can essentially affect the width of the boundary layer near the surface of the squirmer, which in turn affects $V$ and $J$.

\subsection{Effect of multiple stroke modes} \label{sec:r3}

Now that we've gained intuition for how variable viscosity affects treadmill squirmers, we want to examine how the combination of different surface modes affects the swimming and feeding performance. Following previous studies we only consider the first two modes of surface motion because the first few modes have a dominant role in the swimming performance of squirmers \citep{shoele2018effects,michelin2011optimal,magar2003nutrient,pedley2016spherical}. Different combinations of the first (treadmill) and the second (pusher/puller) mode are examined; in the constant-viscosity case, propulsion is entirely generated by the first mode while the second mode only mixes the fluid. In the case of nutrient-dependent viscosity, this mixing changes the local viscosity of the fluid and therefore the second mode can possibly affect the velocity of the squirmer.

\begin{figure}
\centering
\includegraphics[scale=0.8]{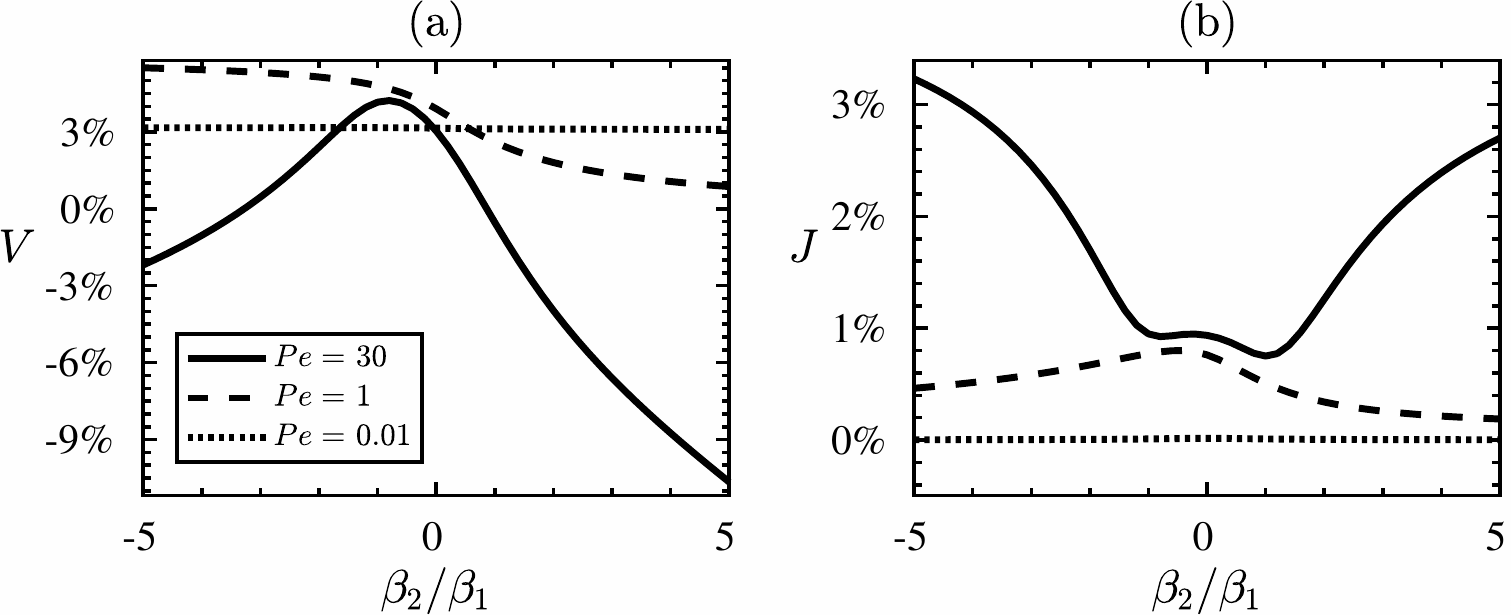}
\caption{Effect of $\beta_2/\beta_1$ ratio on $V$ and $J$ for diffusion-dominated ($\Pen=0.01$), balanced ($\Pen=1$) and advection-dominated ($\Pen=30$) regimes. $V$ is asymmetrically favorable toward negative ratios, while $J$ is reasonably symmetric in $\beta_2/\beta_1$ and increases monotonically with \Pen.}
\label{fig:r3_params}
\end{figure}

In figure \ref{fig:r3_params}(a) we see an interesting asymmetry in $V$ when $\beta_2/\beta_1>0$ (puller-like) vs. when $\beta_2/\beta_1<0$ (pusher-like), as opposed to \ref{fig:r3_params}(b) which is approximately symmetric in $\beta_2/\beta_1$. For low \Pen, all mode combinations have the same effect giving a small boost in swimming speed and do not induce significant changes in nutrient flux compared to the constant-viscosity case. Increasing to $\Pen=1$ there is an equal contribution of advection and diffusion and $\beta_2/\beta_1<0$ has larger $V$ than $\beta_2/\beta_1>0$, although for all $\beta_2/\beta_1$ the positive increase in $V$ has been observed. Nutrient uptake, however, shows a maximum near $\beta_2/\beta_1=0$ indicating that the feeding is optimized for the first mode in this \Pen{} regime.  As \Pen{} is increased even further, the effect of mixing from $\beta_2$ has an adverse effect on $V$ for $\beta_2/\beta_1\gtrapprox 0.5$ and $\beta_2/\beta_1\lessapprox 3$ while $J$ undergoes a rapid growth in these regions.  To summarize, the effect of different $\beta_2/\beta_1$ becomes more prevalent as \Pen{} increases and causes different changes in modifications to velocity than nutrient flux.

\begin{figure}
    \centering
    \includegraphics[scale=0.8]{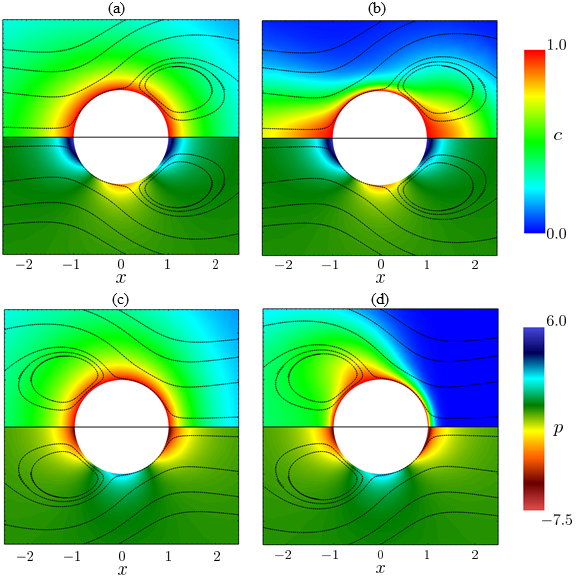}
    \caption{Effect of stroke mode on nutrient/pressure fields. Top (bottom) half of figures corresponds to nutrient (pressure). (a) $\beta_2/\beta_1=-3, \Pen=1$, (b) $\beta_2/\beta_1=-3, \Pen=20$, (c) $\beta_2/\beta_1=3, \Pen=1$, (d) $\beta_2/\beta_1=3, \Pen=20$.}
    \label{fig:r3_flows}
\end{figure}

The $c$ and pressure profiles for four different $\beta_2/\beta_1$-\Pen{} combinations are shown in figure \ref{fig:r3_flows}. Figure \ref{fig:r3_flows}(a,b) shows $\beta_2/\beta_1<0$ (puller-like) for low and high \Pen, respectively. The stagnation point on the surface is placed slightly to the right, and a stagnation point in the flow induces some circulation to the right side of the squirmer. These stagnation points are reversed in the case of $\beta_2/\beta_1>0$ (pusher-like), as shown in figures \ref{fig:r3_flows}(c,d). All swimmers have positive velocity, so the puller-like squirmers induce a stagnation point in front of their direction of motion while the pusher-like squirmers induce a stagnation point behind their direction of motion. Comparing the high-\Pen{} figures \ref{fig:r3_flows}(b,d) it appears that the boundary layer is thin at different places (thin on top for pusher-like and on sides of puller-like) affecting $V$ but the net flux across the boundary is approximately the same for both. This is a possible explanation for why the effect of $V$ was asymmetric in $\beta_2/\beta_1$ but $J$ is roughly symmetric in figure \ref{fig:r3_params}.

\subsection{Effect of Shape} \label{sec:r4}

We now expand our discussion beyond spheres to spheroidal squirmers. As mentioned previously, the surface area is kept constant to compare the effect of geometry on $V$ and $J$. We present the results for different the aspect ratio $\ell$ of the spheroid at the fixed $\Pen=10$, $k=0.5$ for three representative $\beta_2/\beta_1$ ratios selected based on results from section \ref{sec:r3}. 

\begin{figure}
\centering
\includegraphics[scale=0.8]{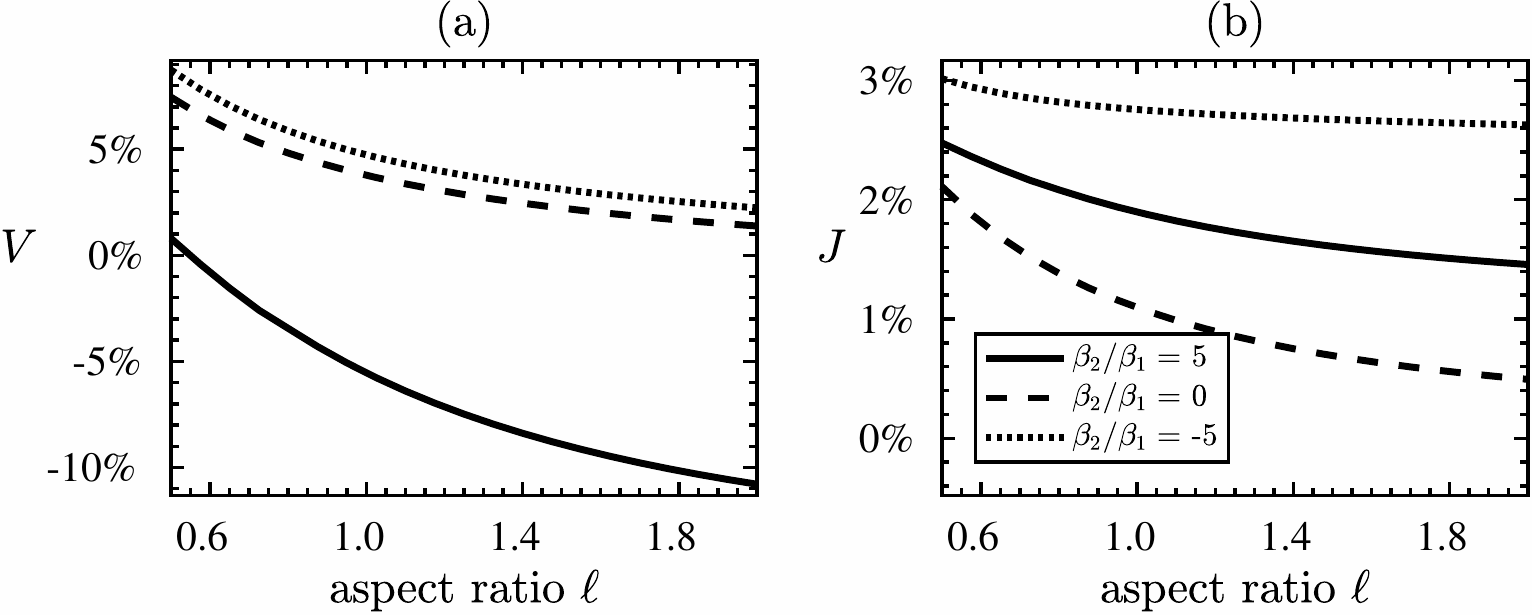}
\caption{Effect of shape on the on $V$ and $J$ for $\beta_2/\beta_1=-5,0,5$.  The oblate and prolate spheroids are shown with $\ell<1$ and $\ell>1$ respectively. All squirmers have positive velocity, so (a,b) pusher-like squirmers induce stagnation points in front, while (c,d) puller-like squirmers induce stagnation points behind. }
\label{fig:r4_params}
\end{figure}

As shown in figure \ref{fig:r4_params}, for all $\beta_2/\beta_1$, both $V$ and $J$ monotonically decrease in $\ell$, meaning that oblate spheroids ($0<\ell<1$) have more positive changes in velocity and feeding due to spatially varying viscosity than prolate spheroids ($\ell>1$) of the  same surface area. Additionally, there is an asymmetry in the $\beta_2/\beta_1$ ratios. In figure \ref{fig:r4_params}(a) $V$ is positive but decreases to zero as $\ell$ increases for $\beta_2/\beta_1\le 0$, while for $\beta_2/\beta_1>0$, $V$ is negative for all but the most oblate spheroid. In figure \ref{fig:r3_params}(b), $J$ is positive, although decreasing monotonically in $\ell$, for all $\beta_2/\beta_1$ ratios, with the highest fluxes experienced for $\beta_2/\beta_1\neq 0$. This suggests that a micro-swimmer could use the mixing mode to increase their nutrient feeding.

\begin{figure}
\centering
\includegraphics[scale=0.6]{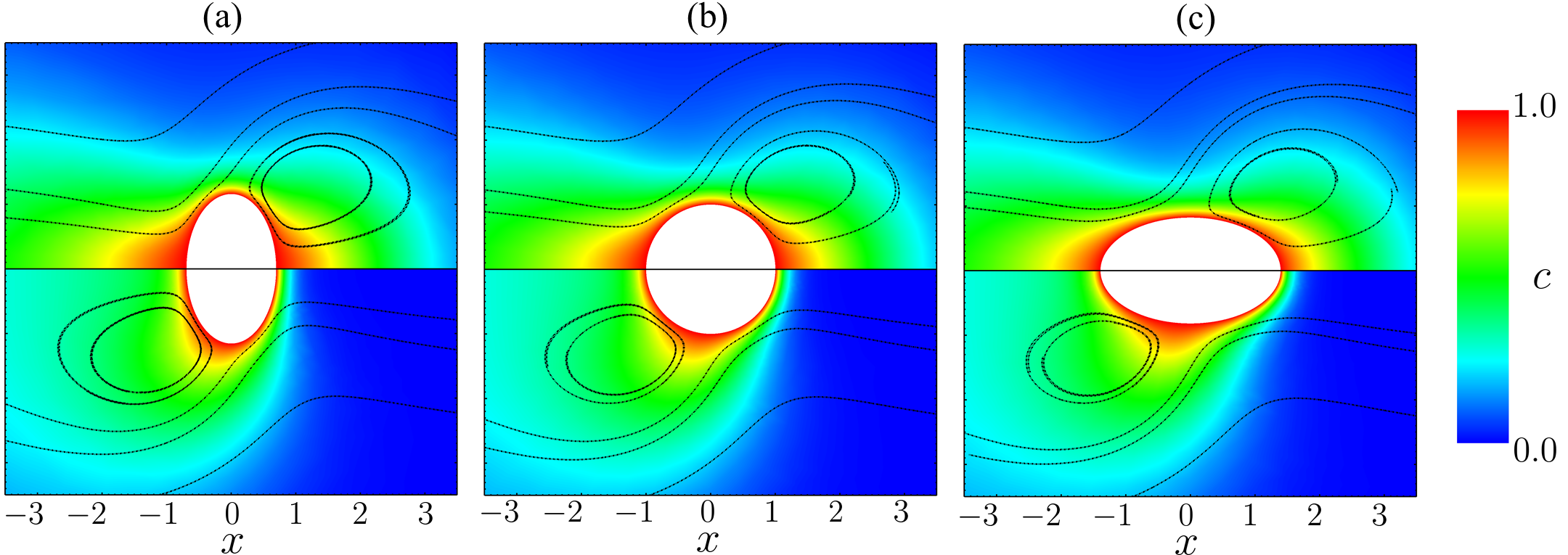}
\caption{Effect of shape on nutrient field. Top panels corresponds to $\beta_2/\beta_1 = -5$, and the bottom panels corresponds to $\beta_2/\beta_1=5$; (a) $\ell=0.6$, top $V = 7.6\%$, $J=2.9\%$, bottom $V=-0.8\%$, $J=2.3\%$, (b) $\ell=1.0$, top $V=4.7\%$, $J=2.8\%$, bottom $V=-5.5\%$, $J=2.7\%$, (c) $\ell=1.7$, top $V=2.7\%$, $J=2.7\%$, bottom $V=-9.8\%$, $J=1.5\%$}
\label{fig:r4_flows}
\end{figure}

Figure \ref{fig:r4_flows} shows three representative spheroids along with their $c$ profile and streamlines for two different surface actuation of $\beta_2/\beta_1=-5$ (top panels) and $\beta_2/\beta_1=5$ (bottom panels). Like for spherical squirmers in section \ref{sec:r3}, all squirmers have positive velocity. Squirmers with $\beta_2/\beta_1=-5$ induce a stagnation point \emph{in front of} the squirmer, while for $\beta_2/\beta_1=5$ a stagnation point is \emph{behind} the squirmer. The stagnation point in front of the squirmer leads to improved swimming performance over the constant-viscosity case, while the stagnation point behind the squirmer has a negative effect on $V$. The flow profiles are qualitatively similar for all $\ell$.

\subsection{Effect of asymmetry of the nutrient distribution on the body} \label{sec:r5}

In \cref{sec:r1,sec:r2,sec:r3,sec:r4} we have investigated the effect of various slip velocities that had some contribution from the $\beta_1$ mode. Now we explore the induced propulsion by the asymmetric scalar BC on the surface of the squirmer when paired with symmetric surface actuation by letting $\beta_2=\pm 1$. In the constant-viscosity case, any type of symmetric actuation of the surface (any actuation modes besides the treadmill mode) always results in zero swimming speed. However, in a variable-viscosity fluid, the asymmetry in $c$ can induce asymmetry in the flow field required for non-zero propulsion. Formally, we now let $\surface_2$ be nonempty, and specifically is the left half of the squirmer, i.e. $\surface_2=\big\{\vx\in\surface \,\big|\, x<0 \big\}$. This means $c$ has homogeneous Dirichlet boundary condition on the right half and a no-flux condition on the left half of the squirmer. 

\begin{figure}
\centering
\includegraphics[scale=1]{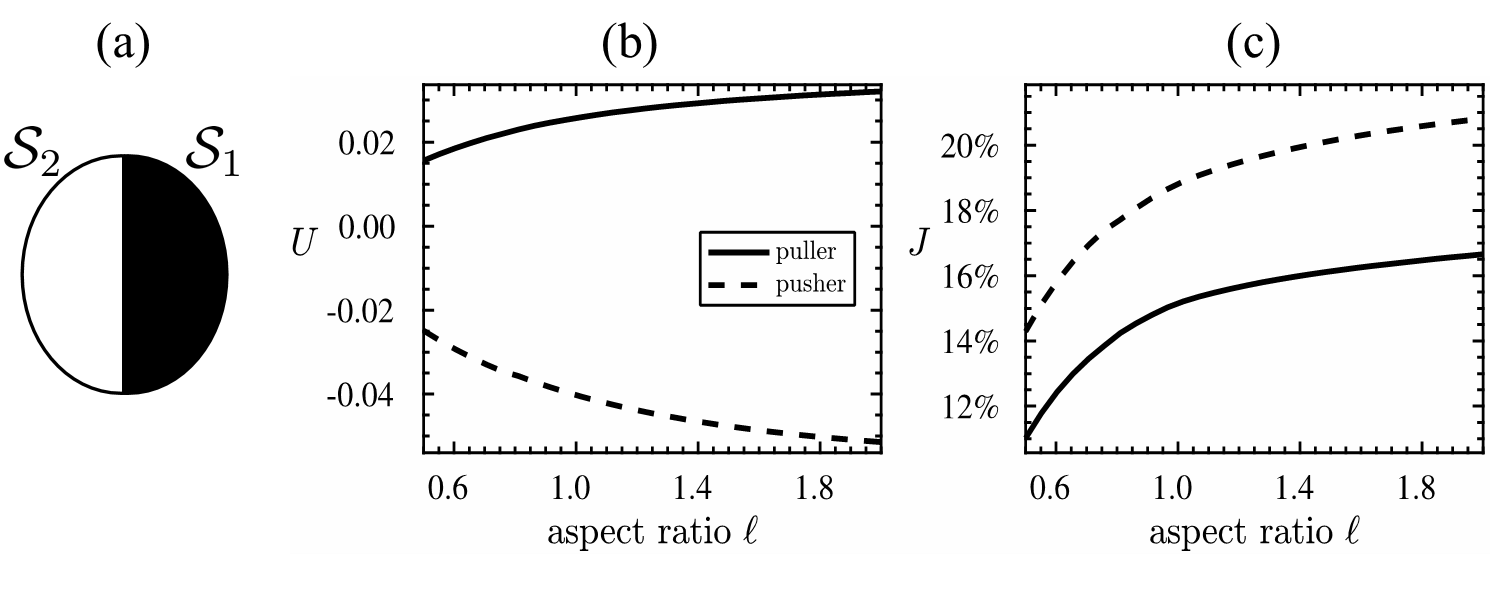}
\caption{Effect of asymmetric nutrient uptake  and symmetric stroke mode ($\beta_2$) on swimming velocity and feeding performance of a spheroidal squirmer. (a) Visual diagram of different surface boundary conditions for nutrient, (b) Non-zero free-swimming velocities and (c) significant increase in nutrient-flux over the constant-viscosity case with same nutrient BC is observed as well.} 
\label{fig:r5_params}
\end{figure}

Figure \ref{fig:r5_params}(b) shows the swimming speed $U$ (not normalized) for pure puller ($\beta_2=1$) and pusher ($\beta_2=-1$) squirmers at $\Pen{}= 10$ when the surface motion only consists of the second mode, $\beta_2$. Free-forced swimming condition for pushers ($\beta_2>0$) requires negative velocity $U$. On the other hand, the puller squirmer ($\beta_2<0$) has zero drag when it has positive velocity $U$. While the relative magnitude of $U$ compared to the values presented in the previous sections is small, the fact that it is finite (nonzero) suggests a new mechanism of propulsion that is fundamentally different from possibilities in constant-viscosity fluids. 

The dependence of $J$ with the aspect ratio is shown in figure \ref{fig:r5_params}(c). It is noted that $J$ increases monotonically with aspect ratio, and $J>0$ for all tested cases of pushers and pullers. The relatively large increase in $J$ indicates that asymmetric $c$ absorption on the surface, paired with a $c$-dependent viscosity, leads to increases in the nutrient flux across the surface of the squirmer. Very interestingly, the increase in nutrient uptake in these cases are much higher than 
\cref{sec:r1,sec:r2,sec:r3,sec:r4}. This suggests that feeding at a localized area -- such as the cell mouth of \emph{Paramecium} \citep{wichterman1986biology} -- is advantageous in an environment with nutrient-dependent viscosity.

\begin{figure}
    \centering
    \includegraphics[scale=0.9]{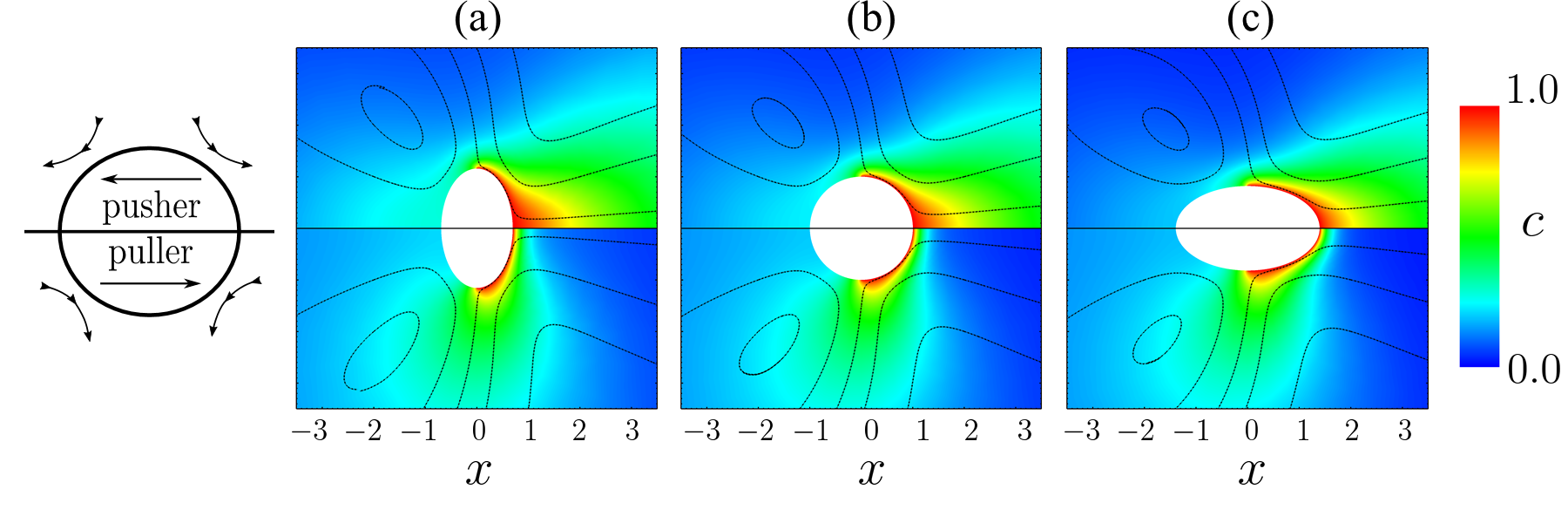}
    \caption{Nutrient distribution corresponding to the asymmetric nutrient uptake at the surface and symmetric swimming stroke modes. Asymmetric nutrient uptake at the surface leads to non-zero propulsion, a major change from the constant-viscosity case. (a) $\ell=0.6$; pusher (top) $U=0.029$, $J=15.8\%$, puller (bottom) $U=-0.018$, $J=12.4\%$, (b) $\ell=1.0$; pusher (top) $U=0.04$, $J=18.8\%$, puller (bottom) $U=-0.03$, $J=15.2\%$, (c) $\ell=1.7$; pusher (top) $U=0.049$, $J=20.5\%$, puller (bottom) $U=-0.03$, $J=16.3$.}
    \label{fig:r5_flows}
\end{figure}

The flow and nutrient fields for three representative aspect ratios of $\ell=0.6, 1.0$ and $1.7$ at $\Pen=10$ are shown in figure  \ref{fig:r5_flows} wherein the top panels are for the pushers ($\beta_2=-1$) and the bottom panels are for the pullers ($\beta_2=1$). In both cases, there are only minor changes in the streamlines, but since the approaching flow for net-zero force swimming squirmers is in different directions, the vector-fields are roughly the negative of each other. The $c$ fields are drastically different due to the different surface velocities, which induce very different viscosity fields and leads to the large differences in $U$ and, to a lesser degree, $J$. 

Comparing figures \ref{fig:r5_params} and \ref{fig:r5_flows}, we see that pushers swim in the opposite direction of the absorption boundary condition on their surface, while pullers swim in the same direction. Therefore it is theoretically possible to obtain both positive and negative propulsion with the same asymmetric nutrient BC simply by switching the sign of the second swimming mode. Equivalently, for a fixed surface velocity, one could change which side of the squirmer absorbed nutrients to change the direction of propulsion. This provides an interesting and currently unprecedented control pathway that can be incorporated on micro-robots.

\section{Conclusions} \label{sec:conclusion}
In this paper we discuss how a variable-viscosity environment influences the swimming and feeding performances of a spheroidal squirmer. Environments with variable viscosity are found to be an influencing factor in the swimming speed and diffusive-flux, represented by their nondimensionalized quantities $V$ and $J$, respectively. The effects of nutrient-induced variable viscosity depend on the relative strengths of advection to diffusion (represented through  \Pen), the type of $c$-dependent viscosity relationship  (through $k$ and $\xi$), stroke mode coupling (through $\beta_2/\beta_1$), the shape of the spheroid (through aspect ratio $\ell$) and where nutrient is absorbed on the surface (through $\surface_2$). We observed that all of the above parameters had some effect on $V$ and $J$, and more interestingly, it is found that an asymmetric boundary conditions of $c$ could induce propulsion for the case which are non-motile in constant-viscosity fluids.

The results presented in section \ref{sec:r5} have will be the most useful to the artificial micro-swimmer community. For example, if one considers the scalar field $c$ to instead represent temperature, then our findings suggest that, when paired with symmetric slip velocities, asymmetrically cooling one side of the squirmer can produce non-zero propulsion. While the phoretic transport of asymmetrically heated particles has been studied \citep{jiang2010active,bickel2013flow,oppenheimer2016motion}, to our knowledge this is the first example of this effect being demonstrated with nontrivial symmetric slip boundary conditions.

In conclusion, we presented how the locomotion in a fluid with variable viscosity compares to its constant-viscosity counterpart. We showed that variable-viscosity environments with symmetric boundary conditions can still lead to asymmetric thrust and therefore propulsion, as long as the scalar field in question contains an asymmetric nature. These results have implications for designing the next generation of artificial micro-swimmers.

\bibliographystyle{jfm}
\bibliography{squirmer}

\end{document}